\documentclass[a4paper,11pt]{article}
\usepackage{amsmath,amssymb,color,comment}
\usepackage[normalem]{ulem}
\usepackage{cancel}
\usepackage{slashed, tensor, bm, physics}
\usepackage{caption}
\usepackage{graphicx}
\usepackage{multirow}
\usepackage{float}
\usepackage[compat=1.1.0]{tikz-feynhand}
\usepackage{jheppub} 

\usepackage{booktabs} 

\newcommand{\SU}{{\rm SU}}

\title{Sommerfeld enhancement from unstable final-state particles in dark matter annihilation}

\author[a]{Tomohiro Abe,}
\author[b]{Ryosuke Sato,}
\author[b]{and Takumu Yamanaka}

\affiliation[a\,]{Department of Physics and Astronomy, Faculty of Science and Technology, Tokyo University of Science, Yamazaki, Noda, Chiba 278-8510, Japan}
\affiliation[b\,]{Department of Physics, The University of Osaka, Machikaneyama-cho, Toyonaka, Osaka 560-0043, Japan}

\emailAdd{abe.tomohiro@rs.tus.ac.jp}
\emailAdd{rsato@het.phys.sci.osaka-u.ac.jp}
\emailAdd{yamanaka@het.phys.sci.osaka-u.ac.jp}

\abstract{
  We study the Sommerfeld enhancement of the annihilation cross section of dark matter into heavier unstable particles. In this process, the annihilation products become non-relativistic near the kinematical threshold. If they experience long-range interactions with each other, their wave function is distorted from a plane wave, and the annihilation cross section can be significantly enhanced. When evaluating the Sommerfeld enhancement from the long-range interactions between the annihilation products, the decay of the products needs to be taken into account.
  We treat this issue by including the decay width in the Schr\"odinger equations of the two-body wave function of the annihilation products. We find that bound states of the annihilation products with a narrow decay width enhance the annihilation cross section through a resonant effect. At the same time, this formulation automatically includes the annihilation process with off-shell final-state particles, which is relevant for a wide decay width. 
  We show that the resonant effect significantly affects the prediction of the dark matter relic abundance.
}

\begin{document} 
\begin{flushright}
OU-HET-1298
\end{flushright}
\maketitle
\flushbottom

\section{Introduction}
The nature of dark matter (DM) remains one of the central open problems in both cosmology and particle physics. Although extensive astrophysical and cosmological observations indicate that about one quarter of the total energy density of the present universe is composed of DM~\cite{Planck:2018vyg}, its microscopic properties are still unknown. A well-motivated possibility is that DM consists of a new stable particle arising from an extension of the Standard Model (SM), whose energy density can be naturally explained by the thermal freeze-out mechanism. In this framework, the present DM relic abundance is determined by its annihilation cross section.

If a long-range interaction exists between non-relativistic DM particles, their two-body wave function is distorted from a plane wave, leading to a significant enhancement of the annihilation cross section. 
This non-perturbative effect is known as the Sommerfeld enhancement (SE)~\cite{Sommerfeld:1931qaf, Sakharov:1948plh}. In the context of DM phenomenology, the SE was originally applied to the heavy DM in an additional multiplet of the SM $\SU (2)_L$ gauge group~\cite{Hisano:2003ec,Hisano:2004ds,Cirelli:2007xd}. Subsequent studies have analyzed its implications in more general DM models (e.g.~\cite{Arkani-Hamed:2008hhe,Iengo:2009ni,Cassel:2009wt,Slatyer:2009vg,Blum:2016nrz,Parikh:2024mwa}).

Most of the previous studies of the SE have focused on the long-range interactions in the initial state, either between DM particles or their co-annihilators.
However, it has recently been pointed out that the SE can also arise from long-range interactions between the annihilation products of DM~\cite{Cui:2020ppc,deLima:2022joz}. 
For instance, in forbidden channels~\cite{Griest:1990kh,DAgnolo:2015ujb}, the kinetic energy of the final-state particle is smaller than that of the initial state particle because the final-state particle is heavier than the initial state particle. As a result, the velocity of the annihilation products is smaller than that of the DM particles. It is known that the SE becomes sizable at small velocities. Therefore, in forbidden channels, the SE in the final state is expected to be more significant than that in the initial state if the final-state potential is the same or stronger compared to that for the initial state. 

In the forbidden DM scenario,
the annihilation products must decay into particles in the thermal bath. 
If their lifetime is longer than the typical timescale of the long-range interactions, the SE can give sizable corrections. On the other hand, for annihilation products with a short lifetime, long-range potential effects are expected to be negligible. Therefore, the SE factor in the final state strongly depends on the lifetime of the final-state particle, and this lifetime must be properly taken into account in the calculations. In the previous work~\cite{Cui:2020ppc}, the effect of the lifetime is included as a cutoff on the velocity of annihilation products, $v_\text{cut}= \sqrt{\Gamma/m_2}$, where $m_2$ is the mass of the annihilation product. The SE is included only if the velocity is larger than $v_\text{cut}$, and is neglected near the kinematical threshold. 
The effects of the bound state were also studied in ref.~\cite{Wang:2022avs}.

However, as shown in studies of top-antitop ($t\bar t$) pair production at collider experiments \cite{Fadin:1987wz,Strassler:1990nw,Sumino:1992ai,Hoang:2000yr,Hagiwara:2008df}, the SE is indeed effective near the threshold. Furthermore, it has a significant effect even below the threshold because of resonant effects of the bound states. 
The $t\bar t$ pair production cross section near the threshold is formulated using the Green's function of non-relativistic final-state particles, and the effect of the decay width is also consistently taken into account. 
Similarly to the $t\bar t$ pair production, bound states of annihilation products may exist in forbidden DM scenarios, and the annihilation cross section in the threshold region can be significantly modified by those bound states. 

In this work, we investigate the final-state SE in DM annihilation into heavier unstable particles.
We formulate the DM annihilation process using the Schr\"odinger equation including the decay width of the annihilation products, and solve it at the leading order of the short-range interaction.
Our formulation is equivalent to the formulation of production cross sections of $t\bar t$~\cite{Fadin:1987wz,Strassler:1990nw,Sumino:1992ai,Hoang:2000yr,Hagiwara:2008df} and $W^+W^-$~\cite{Fadin:1993kg,Fadin:1993kt,Bardin:1993mc,Fadin:1994pm}. 
In this formulation, both of the bound state effect and the contribution of annihilation processes with off-shell final-state particles are automatically included. 
We find that, for a narrow decay width, the bound states of the annihilation products 
can enhance the annihilation cross section through the resonant effect. 
We also show that 
the annihilation processes with off-shell final-state particles are sizable for a large decay width. 
We show that the resonant effect has an impact on the predictions of the relic density in the framework of forbidden DM. 

The remainder of this paper is organized as follows. In section \ref{sec:sommerfeld}, we formulate the final-state SE, and compare it with the formulation given in ref.~\cite{Cui:2020ppc}. In section~\ref{sec:abundance}, we review the formulation of thermally averaged cross sections of forbidden channels and the relic abundance. In section~\ref{sec:examples}, we show two examples of the final-state SE under the attractive Coulomb potential and the attractive Hulth\'en potential. Section~\ref{sec:conclusion} is devoted to the conclusion.

\section{Sommerfeld enhancement in the two-body final state}\label{sec:sommerfeld}
In this section, we formulate the annihilation process of DM including the final-state SE, which is analogous to the $t\bar t$ production at the kinematical threshold \cite{Fadin:1987wz,Strassler:1990nw,Sumino:1992ai,Hoang:2000yr,Hagiwara:2008df}. We also briefly review the formulation of ref.~\cite{Cui:2020ppc} to be compared with our formulation in section \ref{sec:examples}.

\subsection{Enhancement factor with decay width of annihilation products}
We consider a two-to-two annihilation process of DM $\chi_1$ and its anti-particle $\bar{\chi}_1$ with mass $m_1$ into a heavier unstable particle $\chi_2$ and its anti-particle $\bar{\chi}_2$ with mass $m_2\, (> m_1)$. We assume that $\chi_2$ and $\bar{\chi}_2$ decay into some particles in the thermal plasma with the decay width $\Gamma$.
Throughout our discussion, we focus on the $s$-wave annihilation processes.
We assume that both the DM and the annihilation products are non-relativistic, and that only $\chi_2$ and $\bar\chi_2$ have a long-range interaction among them. 
We also suppose that both $\chi_1$ and $\chi_2$ are scalar particles and do not consider spin degrees of freedom for simplicity.

The SE factor is obtained by solving the following Schr\"odinger equations of the initial (final) two-body wave function $\Psi_{1}(\bm{r})$ ($\Psi_2(\bm{r})$), with a long-range potential $V(r)$ and a complex short-range potential $u\delta^3(\bm{r})$: 
\begin{align}
  \left[-\frac{1}{2\mu_1}\nabla^2-E\right]\Psi_1(\bm{r})&=u\delta^3(\bm{r})\Psi_2(\bm{0}),\label{eq:schini}\\
  \left[-\frac{1}{2\mu_2}\nabla^2+V(r)-(E-2\delta m+i\Gamma)\right]\Psi_2(\bm{r})&=u^*\delta^3(\bm{r})\Psi_1(\bm{0}),\label{eq:schfin}
\end{align}
where $r=|\bm{r}|$, $\mu_1=m_1/2$ ($\mu_2=m_2/2$) is the reduced mass of the initial (final) two-body state, and
\begin{align}
   \delta m = m_2-m_1.
\end{align}
Note that the complex phase of $u$ can be absorbed by redefinition of the wave functions $\Psi_1$ and $\Psi_2$.
Focusing on the transition from $\chi_1\bar{\chi}_1$ to $\chi_2\bar{\chi}_2$, the asymptotic form of $\Psi_1(\bm{r})$ and $\Psi_2(\bm{r})$ is given by
\begin{align}
\Psi_1(\bm{r})&\to e^{ip_1z}+f_1(\theta)\frac{e^{ip_1r}}{r},\label{eq:initial asympt}\\
\Psi_2(\bm{r})&\to f_2(\theta)\frac{e^{ip_2r}}{r},\label{eq:final asympt}
\end{align}
where $\theta$ is the angle between the vector $\bm{r}$ and the $z$-axis. 
Here, $p_1$ and $p_2$ are defined as
\begin{align}
p_1&\equiv \sqrt{2\mu_1E},\\
p_2&\equiv \sqrt{2\mu_2(E_2+i\Gamma)},
\end{align}
and
\begin{align}
    E_2 \equiv E-2\delta m.\label{eq:finalenergy}
\end{align}
The first term in eq.~(\ref{eq:initial asympt}) is the incoming plane wave of $\chi_1$ along the $z$-axis. We impose that the incoming wave consists only of the $\chi_1\bar{\chi}_1$ two-body state. The outgoing wave of $\chi_2\bar{\chi}_2$ emerges from the short-range potential $u\delta^3(\bm{r})$, and it exponentially damps at large $r$ because of ${\rm Im}p_2 > 0$.
Under these circumstances, $E$ is always positive, whereas $E_2$ can be negative. Therefore, $\chi_2 \bar{\chi}_2$ forms a bound state for $E_2 <0$, or $E < 2 \delta m$. As we will see in section~\ref{sec:examples}, the bound states give sizable contributions to the annihilation cross section. 

Let us introduce the probability current of $\chi_1$ defined by
\begin{align}
  \bm{j}_1(\bm{r})&\equiv\frac{1}{\mu_1}\Im[\Psi_1^*(\bm{r})\nabla\Psi_1(\bm{r})].
\end{align}
The divergence of the current is obtained from eq.~(\ref{eq:schini}) as
\begin{align}\nabla\cdot\bm{j}_1(\bm{r})&=-2\delta^3(\bm{r})\Im(u\Psi_1^*(\bm{0})\Psi_2(\bm{0})),\label{eq:div1}
\end{align}
The above equation shows that $\bm{j}_1$ is not conserved at the origin due to annihilation. The annihilation rate per unit time $\dd P/\dd t$ is given by 
\begin{align}
    \dv{P}{t}=-\int \dd ^3r\nabla\cdot \bm{j}_1(\bm{r}).
\end{align}
The annihilation cross section is defined as the proportionality constant relating the annihilation rate per unit time to the incoming flux, $j_{1}^{\rm in}=p_1/\mu_1\equiv v_{\rm rel}$
as $\sigma j_{\rm in}={\rm d}P/{\rm d}t$. Thus, we obtain
\begin{align}
  \sigma v_{\rm rel} = 2\Im(u\Psi_1^*(\bm{0})\Psi_2(\bm{0})).\label{eq:cs}
\end{align}

In order to obtain $\Psi_2(\bm{0})$ in eq.~(\ref{eq:cs}), we regard $u$ as a perturbation, and solve the Schr\"odinger equation. By using the Born approximation,\footnote{The Born approximation is not applicable if $|u|$ is too large to be regarded as a perturbation. For large $|u|$, we need to evaluate the Green's function including the short-range effect instead of solving eq.~(\ref{eq:green})~\cite{Blum:2016nrz, Flores:2024sfy, Parikh:2024mwa}.} eq.~(\ref{eq:schfin}) can be solved by introducing a Green's function such that 
\begin{align}
  \left[-\frac{1}{2\mu_2}\nabla^2+V(r)-(E_2+i\Gamma)\right]G_2(\bm{r};E_2+i\Gamma)&=\frac{1}{2\mu_2}\delta^3(\bm{r}).\label{eq:green}
\end{align}
The boundary condition of $G_2(\bm{r};E_2+i\Gamma)$ at $r\to \infty$ is chosen to give the outgoing damping spherical wave:
\begin{align}
  G_2(\bm{r};E_2+i\Gamma)\to\frac{d_{p_2}}{4\pi r}e^{ip_2r},\label{eq:boundgreen}
\end{align}
where $d_{p_2}$ is a function of $p_2$, and the normalization of $d_{p_2}$ is determined by eq.~\eqref{eq:green}. In particular, $d_{p_2}=1$ if $V(r)=0$. 
Then, we obtain the wave function $\Psi_2(\bm{r})$ as
\begin{align}
  \Psi_2(\bm{r})=2\mu_2 u^* G_2(\bm{r};E_2+i\Gamma)\Psi_1(\bm{0}).
\end{align}
Substituting into eq.~(\ref{eq:cs}), we find
\begin{align}
  \sigma v_{\rm rel}
  &=4\mu_2|u|^2|\Psi_1(\bm{0})|^2\Im G_2(\bm{0};E_2+i\Gamma).\label{eq:sigv}
\end{align}
Since the initial two-body state $\chi_1\bar{\chi}_1$  is a plane wave up to ${\cal O}(u^0)$, we put $|\Psi_1(\bm{0})|^2=1$.\footnote{If the initial two particles have a long-range interaction, $|\Psi_1(\bm{0})|^2$ is also modified because of the initial state SE, which is evaluated in the conventional way discussed in refs.~\cite{Hisano:2003ec,Hisano:2004ds,Cirelli:2007xd,Arkani-Hamed:2008hhe,Iengo:2009ni,Cassel:2009wt,Slatyer:2009vg,Blum:2016nrz,Parikh:2024mwa}.}
Note that the effect of the long-range potential $V(r)$ is embedded only in $\Im G_2(\bm{0};E_2+i\Gamma)$.

In order to evaluate the Green's function, it is useful to introduce the function $g_2(r;E_2+i\Gamma)$ defined by
\begin{align}
    G_2(\bm{r};E_2+i\Gamma)=\frac{g_2(r;E_2+i\Gamma)}{4\pi r}.
\end{align}
From eqs.~(\ref{eq:green}) and (\ref{eq:boundgreen}), we obtain the following equation of $g_2(r;E_2+i\Gamma)$:
\begin{align}
    \left[-\frac{1}{2\mu_2}\dv[2]{r}+V(r)-(E_2+i\Gamma)\right]g_2(r;E_2+i\Gamma)&=0,\label{eq:reducegreeneq}
\end{align}
where the boundary conditions for $g_2$ are given as
\begin{align}
    g_2(0;E_2+i\Gamma)&=1,\label{eq:reducegreenbc1}\\
    \lim_{r\to\infty}g_2(r;E_2+i\Gamma)&=d_{p_2}e^{ip_2r}\label{eq:reducegreenbc2}.
\end{align}
We are interested in the behavior of $g_2$ at $r\to0$. We expand the function around $r=0$ as
\begin{align}
    g_2(r;E_2+i\Gamma)=1+rg_2'(0;E_2+i\Gamma)+\cdots
\end{align}
where the prime is the derivative with respect to $r$. Note that there is a term proportional to $r\log r$ if $V(r)$ involves a $1/r$ term around $r\simeq 0$. However, it contributes only to $\Re G_2(\bm{0};E_2+i\Gamma)$, and it does not affect the annihilation cross section which is proportional to $\Im G_2(\bm{0};E_2+i\Gamma)$.
Then, we can rewrite $\Im G_2(\bm{0};E_2+i\Gamma)$ as
\begin{align}
    \Im G_2(\bm{0};E_2+i\Gamma)=\frac{1}{4\pi}\Im g_2'(0;E_2+i\Gamma).
\end{align}
In particular, in the case without the long-range potential, 
\begin{align}
    \Im G_2^{\rm free}(\bm{0};E_2+i\Gamma) &\equiv
    \frac{1}{4\pi}\Re p_2\notag\\
    &=\frac{1}{4\pi}\sqrt{\mu_2\left(\sqrt{E_2^2+\Gamma^2}+E_2\right)}.
    \label{eq:sigvwose}
\end{align}
Using eqs.~(\ref{eq:sigv}) and (\ref{eq:sigvwose}), we obtain the velocity-weighted annihilation cross section of the $s$-wave process without the final-state SE as
\begin{align}
    (\sigma v_{\rm rel})_{\rm w/o\,SE}
    = a\tilde{v}_2,\label{eq:withoutSE}
\end{align}
where $a$ and $\tilde v_2$ are defined as
\begin{align}
    a &\equiv \frac{m_2^2|u|^2}{2\pi},\\
    \tilde{v}_2&\equiv\frac{\Re p_2}{m_2}.
\end{align}
In the following calculation, we parametrize the energy of the initial two-body state in terms of $v_{\rm rel}$ as
\begin{align}
    E=\frac{1}{4}m_1v_{\rm rel}^2.
\end{align}
Then, $\tilde{v}_2$ is expressed as a function of $v_{\rm rel}$:
\begin{align}
    \tilde{v}_2=\Re\sqrt{\frac{m_1v_{\rm rel}^2/4-2\delta m+i\Gamma}{m_2}}.
\end{align} 
Note that $\tilde{v}_2$ is identical to the physical velocity $v_2$ of $\chi_2$ in the limit of $\Gamma\to0$.

The final-state SE factor $S_f(E_2,\Gamma)$ is defined as the ratio of the annihilation cross section with and without the long-range potential.
By using eq.~\eqref{eq:sigv}, $S_f(E_2,\Gamma)$ can be written as the ratio of $\Im G_2(\bm{0};E_2+i\Gamma)$ and $\Im G_2^{\rm free}(\bm{0};E_2+i\Gamma)$ as
\begin{align}
    S_f(E_2,\Gamma) = \frac{\Im G_2(\bm{0};E_2+i\Gamma)}{\Im G_2^{\rm free}(\bm{0};E_2+i\Gamma)}=\frac{\Im g_2'(0;E_2+i\Gamma)}{\Re p_2}.\label{eq:ourSE}
\end{align}
By the definition, $S_f(E_2,\Gamma) = 1$ is satisfied for the case without a long-range potential. Equivalently, the annihilation cross section including the final-state SE is expressed as
\begin{align}
    \sigma v_{\rm rel}=a\tilde{v}_2S_f(E_2,\Gamma).\label{eq:withSE}
\end{align}

Finally, we briefly comment on the consistency of our formulation. Here we utilize the Born approximation for the contact interaction $u$ and treat the decay width $\Gamma$ as an independent parameter. This treatment is consistent only if the annihilation rate of $\chi_2\bar\chi_2$ state into $\chi_1\bar\chi_1$ is much smaller than the total decay width $\Gamma$.
In particular, as we will see in section \ref{sec:examples}, the existence of bound states of $\chi_2 \bar\chi_2$ significantly affects $\chi_1\bar\chi_1$ annihilation. In such cases, the relevant annihilation rate is the partial decay rate of the bound state into $\chi_1 \bar\chi_1$, which is given by
\begin{align}
    \Gamma_{2\to 1} = \frac{\mu_1 p_1 |u|^2}{\pi} |\Psi_2(\bm{0})|^2. \label{eq:gamma bound state}
\end{align}
Here $\Psi_2(\bm{0})$ is a wave function of $\chi_2\bar\chi_2$ state, which is normalized as $\int \dd^3 \bm{r} |\Psi_2(\bm{r})|^2 = 1$. Thus, $\Gamma \gg \Gamma_{2\to 1}$ should be satisfied for the consistency of our formulation.
In section \ref{sec:examples}, we discuss this consistency condition in explicit examples.

\subsection{Comparison with previous work}
The final-state SE has also been studied in ref.~\cite{Cui:2020ppc}, and let us briefly review their formulation.
Instead of including $\Gamma$ in the SE factor, they treated the instability of annihilation products by introducing the cutoff velocity of annihilation products below which the final-state SE is ineffective. In ref.~\cite{Cui:2020ppc}, the final-state SE factor is given by
\begin{align}
    S_f^{\rm (cut)}(E_2)\equiv
    \begin{cases}
        S_f(E_2,0)& (v_{\rm cut}\leq v_2)\\
        1 &(0<v_2<v_{\rm cut})
    \end{cases},\label{eq:luoSE}
\end{align}
where $v_2$ is the velocity of each annihilation product in the center of mass frame. The cutoff velocity $v_{\rm cut}\equiv \sqrt{\Gamma/m_2}$ is obtained by comparing the typical timescale of long-range interaction $(m_2v_2^2)^{-1}$ and the lifetime of the annihilation product $\Gamma^{-1}$. The final-state SE is assumed to be ineffective when the lifetime is shorter than the timescale of long-range interaction: $1 / m_2v_2^2 > 1/\Gamma$.
This condition is equivalent to a condition on $v_2$:
\begin{align}
v_2 <\sqrt{\frac{\Gamma}{m_2}}\equiv v_{\rm cut}.
    \label{eq:def-of-vcut}
\end{align}
Note that $\Gamma$ is negligible in eq.~\eqref{eq:reducegreeneq} if $E_2 \gg \Gamma$. This implies that the SE is relevant to be considered for $E_2 \gtrsim \Gamma$, which also leads to eq.~\eqref{eq:def-of-vcut} up to a numerical coefficient. 
In the following, we refer to this calculation method for the final-state SE as \emph{the cutoff method}. The criterion in the cutoff method is motivated by the Coulomb effects in $W^+W^-$ production near the threshold (see e.g. ref.~\cite{Fadin:1994pm}). We revisit the validity of this criterion in the presence of a long-range potential in section~\ref{sec:examples}. The annihilation cross section with the final-state SE obtained by the cutoff method is evaluated as 
\begin{align}
    \sigma v_{\rm rel}^{\rm (cut)}= a v_2S_f^{\rm (cut)}(E_2).
\end{align}
Note that $\tilde{v}_2$ in eq.~(\ref{eq:withoutSE}) is replaced by $v_2$ in the method. As we will see in section~\ref{sec:examples}, this replacement forbids the non-zero annihilation cross section below the kinematical threshold in the cutoff method.
\section{DM Relic abundance}\label{sec:abundance}

We now discuss the DM relic abundance. We assume that there is no asymmetry between the number densities of $\chi_1$ and $\bar{\chi}_1$. The time evolution of the number density $n_1$ of $\chi_1$ is obtained by solving the following Boltzmann equation:
\begin{align}
    \dv{n_1}{t}+3H n_1&=-\langle\sigma v_{\rm rel}\rangle (n_1^2-n_{1,eq}^2),
\end{align}
where $H$ is the Hubble parameter, and $n_{1,\,eq}$ is the  number density of $\chi_1$ in thermal equilibrium:
\begin{align}
    n_{1,eq}=\left(\frac{m_1T}{2\pi}\right)^{\frac{3}{2}}\exp(-\frac{m_1}{T}),
\end{align}
where $T$ is the temperature of the thermal bath. 
We assume that $\chi_2$ and $\bar{\chi}_2$ remain in thermal equilibrium through their decay and inverse decay processes with the thermal plasma. 
Defining the yield $Y\equiv n_1/s$, where $s$ is the entropy density, we obtain 
\begin{align}
    \dv{Y}{x}=-\frac{\langle \sigma v_{\rm rel} \rangle}{Hx}\left(1-\frac{x}{3g_{*s}}\dv{g_{*s}}{x}\right)s\left(Y^2-Y_{eq}^2\right),\label{eq:boltzmann}
\end{align}
where $x=m_1/T$. The quantity $Y_{eq}=n_{1,eq}/s$ is the yield of $\chi_1$ in thermal equilibrium. When we solve eq.~(\ref{eq:boltzmann}), we take the temperature dependence of $g_{*s}$ into account~\cite{Saikawa:2020swg}. 
Solving the differential equation, we obtain $Y$ as a function of $x$. 
Since both $\chi_1$ and ${\bar\chi}_1$ contribute to the DM energy density,
the present relic abundance $\Omega h^2$ is evaluated using
the present $\chi_1$ yield $Y_0$ as
\begin{align}
    \Omega h^2=\frac{2m_1s_0Y_0}{\rho_{c,0}}h^2\simeq5.49\times10^8~{\rm GeV^{-1}}\times m_1Y_0,
\end{align}
where we have used the present critical density $\rho_{c,0}\simeq 1.053\times 10^{-5} h^2~{\rm GeV\cdot cm^{-3}}$ and the present entropy density  $s_0=2891.2~{\rm cm^{-3}}$~\cite{ParticleDataGroup:2024cfk}. 

We now evaluate the thermally averaged annihilation cross section. 
The thermally averaged annihilation cross section with the final-state SE is given by
\begin{align}
    \langle\sigma v_{\rm rel}\rangle=\frac{a}{2\sqrt{\pi}} x^{\frac{3}{2}}\int_0^\infty \dd{v_{\rm rel}} v_{\rm rel}^2S_f(E_2,\Gamma)\tilde{v}_2\exp(-\frac{v_{\rm rel}^2x}{4}).\label{eq:sigvave}
\end{align}
Note that the thermally averaged annihilation cross section without the final-state SE
is obtained by setting $S_f(E_2,\Gamma)=1$ in eq.~(\ref{eq:sigvave}).

In contrast, in the cutoff method, the annihilation cross section below the kinematical threshold is set to zero. The value of $v_{\rm rel}$ at the kinematical threshold is given by
\begin{align}
     v_{\rm th}=2\sqrt{\left(\frac{m_2}{m_1}\right)^2-1}.
\end{align}
The thermally averaged annihilation cross section in the cutoff method  $\langle\sigma v_{\rm rel}\rangle^{\rm (cut)}$ is obtained by averaging over the velocity above $v_{\rm th}$:
\begin{align}
    \langle\sigma v_{\rm rel}\rangle^{\rm (cut)}&=\frac{a}{2\sqrt{\pi}} x^{\frac{3}{2}}\int_{v_{\rm th}}^\infty \dd v_{\rm rel}\,v_{\rm rel}^2S_f^{\rm (cut)}(E_2)v_2\exp(-\frac{v_{\rm rel}^2x}{4}).\label{eq:sigvavecut}
\end{align}
In the following section, we compare the impact of eqs.~(\ref{eq:sigvave}) and (\ref{eq:sigvavecut}) on the relic abundance.

Before closing this section, we comment on the finite temperature effect on the bound states. As we discuss in the following, the long-range interaction between the annihilation products can form the bound states similarly to the $t\bar{t}$ production near the threshold~\cite{Fadin:1987wz,Strassler:1990nw,Sumino:1992ai,Hoang:2000yr,Hagiwara:2008df}. Finite temperature of the thermal bath induces the thermal mass of the mediators of the long-range interactions $m_{\rm th}^2\sim \alpha T^2$.
Therefore, if the temperature is too high, the long-range force is screened, and the bound states cannot exist.
A conservative criterion for the existence of bound states at finite temperature can be estimated by comparing the temperature of the thermal bath and the inverse Bohr radius of the annihilation products~\cite{Kim:2016kxt}. Specifically, we assume that the bound states exist if
\begin{align}
    T\lesssim\alpha m_2,
\end{align}
where $\alpha$ is the coupling of the long-range interaction among the annihilation products.
Under this assumption,
if the freeze-out temperature is $T_f\sim m_1/30$, the bound states exist at freeze-out, provided that
\begin{align}
    \alpha\gtrsim \frac{1}{30}\times\frac{m_1}{m_2}.\end{align}
Thus, bound states can exist at the freeze-out temperature as long as the long-range interactions between the annihilation products are strong enough. In order to discuss the effects of the bound states at the freeze-out temperature, we focus on the final-state SE with $\alpha\sim{\cal O}(0.1)$.

\section{Examples}\label{sec:examples}
In this section, we apply the formulation developed in section \ref{sec:sommerfeld} to two simple examples of the long-range potential between $\chi_2$ and $\bar\chi_2$: the attractive Coulomb potential and the attractive Hulth\'en potential. 
We show that the bound state spectrum significantly affects the DM annihilation cross section and its relic abundance.

\subsection{Attractive Coulomb potential}
Let us derive the final-state SE factor for the attractive Coulomb potential:
\begin{align}
    V(r) = -\frac{\alpha}{r}.
\end{align}
The solution of eq.~(\ref{eq:reducegreeneq}) satisfying the boundary conditions~(\ref{eq:reducegreenbc1}) and (\ref{eq:reducegreenbc2}) is given by the analytic form:
\begin{align}
    g_2^{\rm (C)}(r;E_2+i\Gamma)
    = \Gamma\left(1-i\frac{\mu_2\alpha}{p_2}\right)W\left(i\frac{\mu_2\alpha}{p_2},\frac{1}{2},-2ip_2 r\right),
\end{align}
where $\Gamma(z)$ and $W(k,\mu,z)$ are the gamma function and the Whittaker function, respectively. Then $\Im g_2'^{\rm (C)}(0;E_2+i\Gamma)$ is given by
\begin{align}
    \Im g_2^{{\prime\rm (C)}}(0;E_2+i\Gamma)=\Re p_2 -2\mu_2\alpha\Im\left(\log(\frac{-ip_2}{\mu_2\alpha})+\psi\left(1-i\frac{\mu_2\alpha}{p_2}\right)\right),\label{eq:greencoulomb}
\end{align}
where $\psi(z)\equiv\Gamma'(z)/\Gamma(z)$ is the digamma function. 
The SE is calculated by substituting eq.~\eqref{eq:greencoulomb} into eq.~\eqref{eq:ourSE}, and we obtain $\sigma v_{\rm rel}$ from eq.~(\ref{eq:withSE}). 
\begin{figure}
        \centering
        \includegraphics[width=0.7\linewidth]{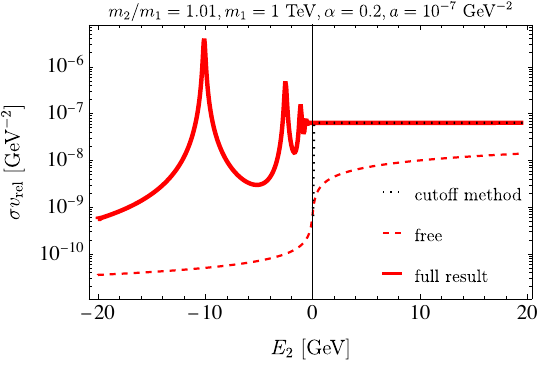}
~\\[5mm]
    \includegraphics[width=0.7\linewidth]{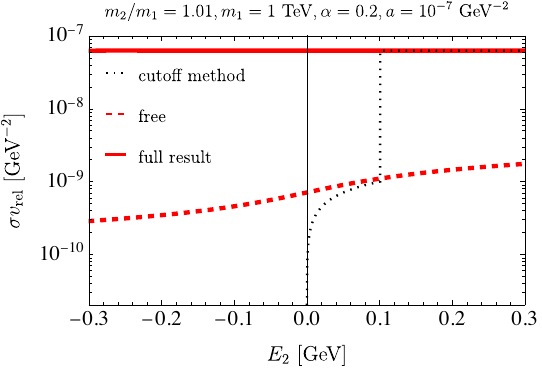}
    \caption{\emph{Top}: Annihilation cross sections of $\chi_1\bar{\chi}_1$ as a function of $E_2$ under the attractive Coulomb potential, $V(r)=-\alpha/r$. We take $m_1=1~{\rm TeV}$, $m_2/m_1=1.01$, $\alpha=0.2$, $a=10^{-7}~{\rm GeV^{-2}}$, and $\Gamma/m_2 = 10^{-4}$. The vertical line at $E_2=0$ indicates the threshold energy of $\chi_1\bar{\chi}_1\to\chi_2\bar{\chi}_2$.
    The black dotted line shows the cross section including the final-state SE obtained by the cutoff method (\ref{eq:luoSE}) with $v_{\rm cut}=10^{-2}$. The red dashed line (\emph{free}) shows the cross section without SE with $\Gamma/m_2=10^{-4}$. The red solid line (\emph{full result}) shows the cross section including the final-state SE defined in eq.~(\ref{eq:ourSE}). \emph{Bottom}: Annihilation cross sections of $\chi_1\bar{\chi}_1$ in the vicinity of $E_2=0$. Parameters and lines are the same as in the top panel.
    }
    \label{fig:integrand}
\end{figure}

First, we discuss the behavior of $\sigma v_{\rm rel}$ as a function of $E_2$.
Figure \ref{fig:integrand} shows $\sigma v_{\rm rel}$ for $m_1=1~{\rm TeV}$, $m_2/m_1=1.01$, $\alpha=0.2$, $a=10^{-7}~{\rm GeV^{-2}}$, and $\Gamma/m_2=10^{-4}$.
For this parameter set, the cutoff velocity for the cutoff method is obtained as $v_{\rm cut}=10^{-2}$ from eq.~\eqref{eq:def-of-vcut}.
We refer to the result of our formulation of the final-state SE as \emph{the full result} and the result without the final-state SE as \emph{the free result} in the following.
If we evaluate the final-state SE with the cutoff method, the annihilation cross section is not enhanced for
$v_2 < v_{\rm cut}$, which corresponds to $E_2\lesssim 0.1~{\rm GeV}$.
On the other hand,  the full result shows that it is indeed enhanced in this energy range and the final-state SE is not negligible. This behavior is clearly shown in the bottom panel of figure~\ref{fig:integrand}. 
Furthermore, including $\Gamma$ in the Green's function, the annihilation cross section becomes non-zero for negative $E_2$, i.e., below the kinematical threshold. This is because annihilation processes which involve an off-shell particle, such as $\chi_1\bar{\chi}_1\to\chi_2\bar{\chi}_2^*$, are kinematically allowed. 
Moreover, as we can see in the top panel, $\sigma v_\text{rel}$ is resonantly enhanced at $E_2\simeq$ $-10.1$ GeV, $-2.52$ GeV, and $-1.12$ GeV, corresponding to binding energies of $1s$, $2s$, and $3s$ bound states of $\chi_2\bar{\chi}_2$, respectively. We can understand this behavior analytically as follows.
In the $u \to 0$ limit, the binding energy ${\cal E}_n$ of the $n$-th $\chi_2 \bar{\chi}_2$ bound state is given by
\begin{align}
    {\cal E}_n=-\frac{\mu_2\alpha^2}{2n^2},\quad (n=1, 2, \cdots). \label{eq:bound state energy coulomb}
\end{align}
Expanding the argument of the digamma function in eq.~\eqref{eq:greencoulomb} around $E_2={\cal E}_n$, we obtain
\begin{align}
    1-i\frac{\mu_2\alpha}{p_2} &= 1-i\frac{\mu_2\alpha}{\sqrt{2\mu_2(E_2+i\Gamma)}} \nonumber\\
    &\simeq 1-n-\frac{n^3}{\mu_2\alpha^2} [ (E_2-{\cal E}_n)+i\Gamma ],\label{eq:argument expansion}
\end{align}
where we have assumed $\Gamma/(\mu_2\alpha^2)\ll1$. 
$\psi(z)$ has a simple pole at $z=-k$, where $k=0,1,2,\cdots$, with residue $-1$. 
Near the pole, its asymptotic behavior is given by $\psi(z)\simeq -1/(z+k)$.
Thus, we obtain
\begin{align}
    \psi\left(1-i\frac{\mu_2\alpha}{p_2}\right)\simeq \frac{\mu_2\alpha^2/n^3}{(E_2-{\cal E}_n)+i\Gamma}.
\end{align}
Its imaginary part induces the Lorentzian function, which is responsible for the resonances:
\begin{align}
    \Im\psi\left(1-i\frac{\mu_2\alpha}{p_2}\right)\simeq-\frac{\mu_2\alpha^2}{n^3}\frac{\Gamma}{(E_2-{\cal E}_n)^2+\Gamma^2}.~\label{eq:lorentzian}
\end{align}
Thus, we obtain $S_f$ around $E_2= {\cal E}_n$ as
\begin{align}
S_f(E_2,\Gamma)\simeq\frac{1}{\Re p_2}\frac{2\mu_2^2\alpha^3}{n^3}\frac{\Gamma}{(E_2-{\cal E}_n)^2+\Gamma^2}.\label{eq:Sfcoulomb}
\end{align}
As we discuss below, the existence of resonances affects the mass ratio of $\chi_1$ and $\chi_2$ that reproduces the observed relic abundance of DM.
\begin{figure}[tbp]
    \centering
    \includegraphics[width=0.7\linewidth]{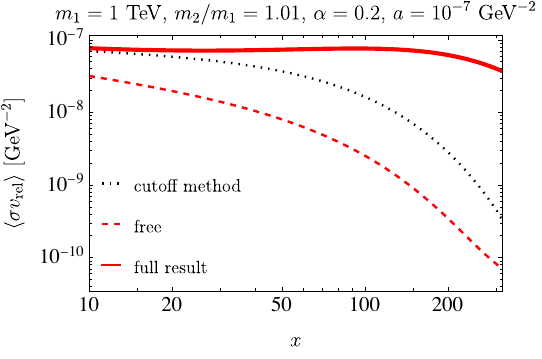}
    \caption{Thermally averaged annihilation cross section as a function of $x=m_1/T$. The parameters are the same as in figure~\ref{fig:integrand}. The red solid line indicates the full result obtained from eq.~(\ref{eq:sigvave}) with $\Gamma/m_2 = 10^{-4}$. The red dashed line indicates the free result obtained from eq.~(\ref{eq:sigvave}) with $S_f(E_2,\Gamma)=1$ and $\Gamma/m_2 = 10^{-4}$. The black dotted line indicates the cutoff method result obtained from eq.~(\ref{eq:sigvavecut}) with $v_{\rm cut}=10^{-2}$.}
    \label{fig:sigmav_ave}
\end{figure}

Next, we discuss the impact of the final-state SE on the thermally averaged annihilation cross section $\langle \sigma v_{\rm rel} \rangle$.
Figure \ref{fig:sigmav_ave} shows the $x=m_1/T$ dependence of the thermally averaged annihilation cross sections given in eqs.~(\ref{eq:sigvave}) and (\ref{eq:sigvavecut}).
We use the same parameter set as in figure~\ref{fig:integrand}. 
For large $x$, the full result differs significantly from the cutoff method result due to the contribution from the resonances of the bound states discussed above. Since the resonances appear below the kinematical threshold, the contribution is significant at low temperature, i.e., at large $x$.

\begin{figure}[tbp]
\centering
    \includegraphics[width=0.65\linewidth]{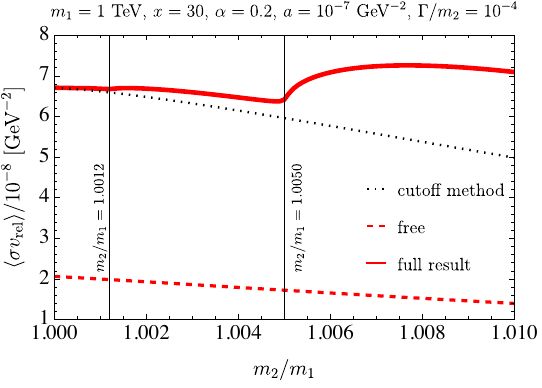}
~\\[3mm]
    \includegraphics[width=0.65\linewidth]{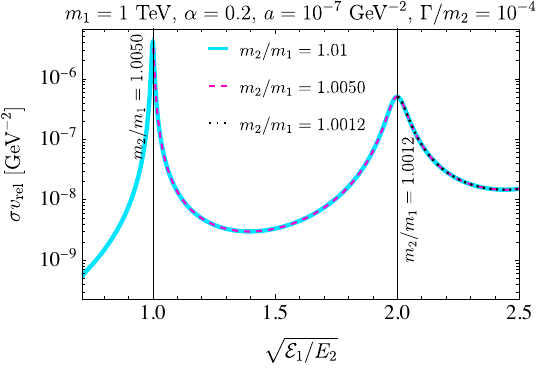}
    \caption{\emph{Top}: Thermally averaged annihilation cross sections as a function of $m_2/m_1$ with $x=30$. We take the same parameters as in figure~\ref{fig:sigmav_ave}. The vertical lines indicate the mass ratios shown in the bottom panel.
    \emph{Bottom}: $\sigma v_{\rm rel}$ as a function of $\sqrt{{\cal E}_1/E_2}$. We focus on the region where $E_2<0$ in this panel. The solid cyan, dashed magenta, and dotted black lines indicate the results with $m_2/m_1=1.01, 1.0050$, and $1.0012$, respectively. The vertical lines indicate the minimum mass ratio for which the $1s$ and $2s$ bound states contribute to the annihilation cross section.
    }
    \label{fig:bound_state_coulomb}
\end{figure}
In the $\chi_1 \bar\chi_1$ annihilation process, the resonance of the $n$-th bound state appears if a kinematical condition $2m_2 + {\cal E}_n \simeq 2m_1 + E \geq 2m_1$ is satisfied. 
Thus, the mass ratio $m_2/m_1$ strongly affects $\langle \sigma v_{\rm rel} \rangle$. The top panel of figure \ref{fig:bound_state_coulomb} shows $\langle \sigma v_{\rm rel} \rangle$ as a function of $m_2/m_1$.
As shown in the top panel, the full result slightly deviates from the cutoff method for $m_2/m_1\gtrsim 1.0012$, and a significant difference between them appears for $m_2/m_1\gtrsim1.0050$. These deviations originate from the contribution of the resonances of the bound states to the integral in eq.~(\ref{eq:sigvave}). 
In the bottom panel of figure \ref{fig:bound_state_coulomb}, we focus on the region with negative $E_2$ and show the cross section $\sigma v_{\rm rel}$ as a function of $\sqrt{{\cal E}_1 / E_2} = \sqrt{\mu_2 \alpha^2/2|E_2|}$.
The spectrum of the bound states given in eq.~\eqref{eq:bound state energy coulomb} tells us that the resonances of the bound states appear if $\sqrt{{\cal E}_1 / E_2}$ is a positive integer. For example, the peaks at $\sqrt{{\cal E}_1/E_2}=1$ and $\sqrt{{\cal E}_1/E_2}=2$ represent the resonances of the $1s$ and $2s$ bound states, respectively.
Note that $E_2$ depends on $v_\text{rel}$ and takes the minimum value at $v_\text{rel} = 0$, 
\begin{align}
E_2^\text{min} = -2 \delta m = -2m_1 \qty(\frac{m_2}{m_1} - 1),
\end{align}
which corresponds to the lower limit of the integral in eq.~\eqref{eq:sigvave}. 
Therefore, the mass ratio $m_2/m_1$ determines whether the annihilation is resonantly enhanced 
by the bound state. 
With the parameter set we use here, we find $E_2^\text{min} \leq \mathcal{E}_2$ for $m_2/m_1 \gtrsim 1.0012$, 
and thus the $2s$ resonance contributes to the annihilation process for $m_2/m_1 \gtrsim 1.0012$.
Similarly, $E_2^\text{min} \leq \mathcal{E}_1$ for $m_2/m_1 \gtrsim 1.0050$,  
and thus the $1s$ resonance contributes to the thermally averaged annihilation cross section for $m_2/m_1>1.0050$.
The vertical lines in the figure correspond to these values.   

The region to the right of each line is kinematically allowed and contributes to the thermally averaged annihilation cross section. 
Comparing the two panels, we find that the contributions from the $1s$ and $2s$ bound states enhance the thermally averaged annihilation cross section.
The kinks in the top panel of figure \ref{fig:bound_state_coulomb} at $m_2/m_1 = 1.0012$ and $1.0050$ correspond to the threshold values of $m_2/m_1$ to include the contributions from the $2s$ and $1s$ bound states, respectively.
Note that the resonances do not appear in the cutoff method because the annihilation cross section is set to zero below the kinematical threshold. 

\begin{figure}[tbp]
    \begin{minipage}[b]{0.49\linewidth}
        \centering
    \includegraphics[width=\linewidth]{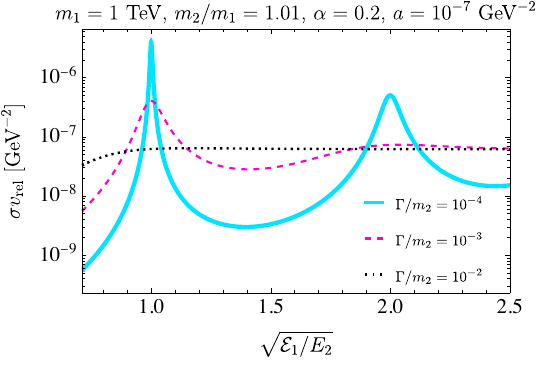}
    \end{minipage}
    \begin{minipage}[b]{0.49\linewidth}
        \centering
        \includegraphics[width=\linewidth]{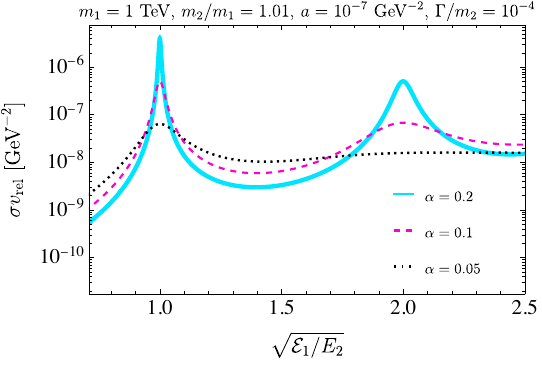}
    \end{minipage}
    \caption{\emph{Left}: $\sigma v_{\rm rel}$ as a function of $\sqrt{{\cal E}_1/E_2}$ for $\Gamma/m_2=10^{-4}$ (cyan solid line), $\Gamma/m_2=10^{-3}$ (magenta dashed line), and $\Gamma/m_2=10^{-2}$ (black dotted line). \emph{Right}: $\sigma v_{\rm rel}$ as a function of $\sqrt{{\cal E}_1/E_2}$ for $\alpha=0.2$ (cyan solid line), $\alpha=0.1$ (magenta dashed line), and $\alpha=0.05$ (black dotted line).}
    \label{fig:sigmav_alpha_gamma}
\end{figure}
We next discuss how the decay width $\Gamma$ and the coupling $\alpha$ affect the predictions for $\sigma v_{\rm rel}$ and $\langle \sigma v_{\rm rel}\rangle$.
Figure~\ref{fig:sigmav_alpha_gamma} shows the dependence of the size of the resonance peak on  $\Gamma$ and $\alpha$. Here, we use the same axes as in the bottom panel of figure~\ref{fig:bound_state_coulomb}. In both panels, we fix the parameters as $m_1=1~{\rm TeV}$, $m_2/m_1=1.01$, and $a=10^{-7}~{\rm GeV^{-2}}$. In the left panel, we fix $\alpha=0.2$, and calculate $\sigma v_{\rm rel}$ in our formulation for $\Gamma/m_2=10^{-4}$, $10^{-3}$, and $10^{-2}$. In the right panel, we fix $\Gamma/m_2=10^{-4}$, while varying $\alpha=0.2, 0.1$, and $0.05$. The resonant behavior of the annihilation cross section around $E_2={\cal E}_n$ is well described by eq.~\eqref{eq:Sfcoulomb} if $\Gamma/(\mu_2\alpha^2)\ll 1$. 
For large $\Gamma/m_2$, the poles of the digamma function in eq.~\eqref{eq:greencoulomb} move away from the real axis, and consequently the resonance peaks at $E_2 = {\cal E}_n$ are suppressed. The peaks are also suppressed for smaller $\alpha$, as can be seen from eq.~\eqref{eq:Sfcoulomb}.

\begin{figure}[tbp]
    \centering
    \begin{minipage}[b]{0.49\linewidth}
        \includegraphics[width=\linewidth]{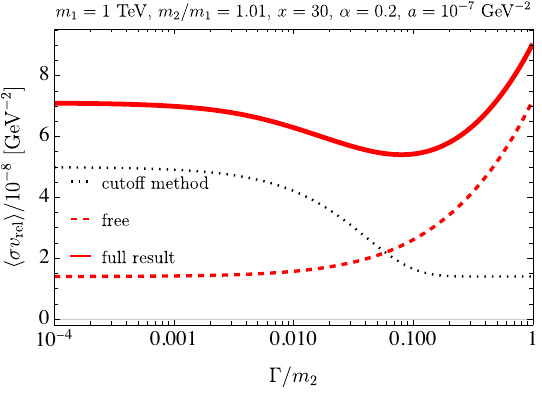}
    \end{minipage}
    \begin{minipage}[b]{0.49\linewidth}
        \includegraphics[width=\linewidth]{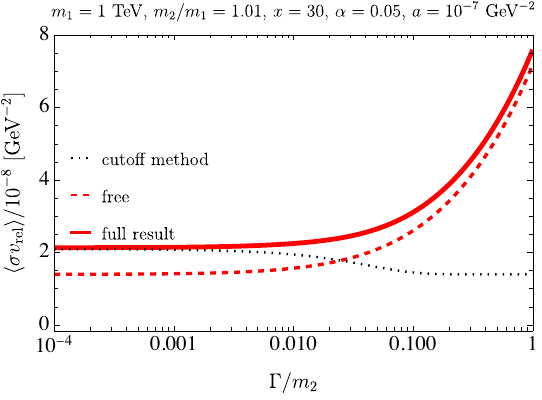}
    \end{minipage}    
    \caption{Thermally averaged annihilation cross section as a function of $\Gamma/m_2$. In both panels, we set $m_1=1~{\rm TeV}$, $m_2/m_1=1.01$, $x=30$, and $a=10^{-7}~{\rm GeV^{-2}}$. In the left panel, we set $\alpha=0.2$, while we set $\alpha=0.05$ in the right panel.}
    \label{fig:sigmav_ave_gamma}
\end{figure}
Figure~\ref{fig:sigmav_ave_gamma} shows the $\Gamma/m_2$ dependence of $\langle\sigma v_{\rm rel}\rangle$. In both panels, we set $x=30$, $m_2/m_1=1.01$, and $a=10^{-7}~{\rm GeV^{-2}}$. The left and right panels show the results for $\alpha=0.2$ and $\alpha=0.05$, respectively. 
If $\Gamma$ is small and $\alpha$ is large, the full result shows that the thermally averaged annihilation cross section is larger than that obtained with the cutoff method as shown in the left panel. This deviation originates from the contribution of the resonances of the bound states. On the other hand, as shown in the right panel, there is little deviation between the two results for small $\alpha$. This is because the annihilation cross section of the full result around $E_2={\cal E}_n$ is proportional to $\alpha^3$,
\begin{align}
    \sigma v_{\rm rel}\simeq a\frac{\mu_2\alpha^3}{n^3}\frac{\Gamma}{(E_2-{\cal E}_n)^2+\Gamma^2},
\end{align}
and the resonance peak is suppressed by small $\alpha$ as discussed above.
For large $\Gamma$, the full result and the free result increase due to the contribution of the processes with off-shell final-state particles.  
On the other hand, the cutoff method does not include off-shell processes, and thus the cross section differs from those obtained by the other two methods. In fact, it reduces to the tree level cross section evaluated with $\Gamma = 0$.
In this regime, the full result approaches the free result, and we can see that the final-state SE becomes ineffective as $\Gamma$ becomes larger.  

Here, we revisit the criterion under which the final-state SE is effective.  
For the Coulomb potential, the deviation of the final-state SE factor from unity is obtained from eqs.~\eqref{eq:ourSE} and \eqref{eq:greencoulomb} as follows:
\begin{equation}
  S_f(E_2,\Gamma) - 1
  = -\frac{2\mu_2\alpha}{\mathrm{Re}\,p_2}\,
  \mathrm{Im}\!\left[
    \log\!\left(\frac{-ip_2}{\mu_2\alpha}\right)
    + \psi\!\left(1 - i\frac{\mu_2\alpha}{p_2}\right)
  \right] .
\end{equation}
Note that the final-state SE factor is determined by
$\mu_2\alpha/p_2$. Using eq.~\eqref{eq:sigvwose} for
$\Re p_2$, we find that
\begin{align}
  \frac{2\mu_2\alpha}{\mathrm{Re}\,p_2}
  =\sqrt{\frac{2m_2\alpha^2}{\sqrt{E_2^2+\Gamma^2}+E_2}}
  \simeq
  \begin{cases}
    \sqrt{\frac{m_2\alpha^2}{E_2}}
      & (E_2 \gg \Gamma),\\[2.2ex]
    \sqrt{\frac{2m_2\alpha^2}{\Gamma}}
      & (|E_2| \ll \Gamma),\\[2.2ex]
    \sqrt{\frac{4m_2\alpha^2}{\Gamma}\,\frac{|E_2|}{\Gamma}}
      & (E_2 < 0,\ |E_2| \gg \Gamma),
  \end{cases}
  \label{eq:Sf-regimes}
\end{align}
where we have rewritten $\mu_2=m_2/2$ in order to compare the criterion of the cutoff method given in eq.~\eqref{eq:def-of-vcut}.
For $E_2 \gg \Gamma$, the final-state particles are on-shell and the decay width is negligible. Since $E_2 \gg \Gamma$ is equivalent to
$m_2 v_2^2 \gg \Gamma$, in this regime the decay width is to be compared with $m_2 v_2^2$, reproducing the intuition behind the cutoff method. For $|E_2| \ll \Gamma$, where the final-state particles can be either on-shell or off-shell, the SE is instead controlled by $m_2\alpha^2/\Gamma$. For $E_2 < 0$ and $|E_2| \gg \Gamma$, the final-state particles are off-shell, and the typical scale of the energy is given by the binding energy, i.e., $|E_2| \sim m_2\alpha^2$. Substituting this value yields $\sqrt{4m_2\alpha^2/\Gamma}\sqrt{|E_2|/\Gamma} \sim m_2\alpha^2/\Gamma$.
We therefore conclude that,  except in the regime where $E_2\gg\Gamma$, the SE remains effective as long as
\begin{align}
    \Gamma\lesssim m_2\alpha^2,
\end{align}
and the full result approaches the free result if $\Gamma$ exceeds $m_2\alpha^2$. This is consistent with the large-width behavior seen in figure~\ref{fig:sigmav_ave_gamma}. As $\Gamma$ increases, the Lorentzian resonances in eq.~\eqref{eq:lorentzian} broaden and their peaks are suppressed, and the full result approaches the free result. The same conclusion follows from the analyses of the Coulomb effects in $W^+W^-$ production near the threshold~\cite{Fadin:1993kg,Bardin:1993mc,Fadin:1993kt,Fadin:1994pm}, where the Coulomb series is reorganized in powers of $\sqrt{m_W\alpha_W^2/\Gamma}$; in our case this expansion parameter reads $\sqrt{m_2\alpha^2/\Gamma}$. 

As we have discussed in section \ref{sec:sommerfeld}, since we use the Born approximation regarding $u$ as the perturbation, our formulation is valid only when 
the annihilation rate of $\chi_2 \bar{\chi}_2$ into $\chi_1\bar{\chi}_1$ is much smaller than the total decay width $\Gamma$. 
We estimate this consistency condition for the $1s$ bound state of $\chi_2\bar\chi_2$ 
because the $1s$ bound state has an important impact on the annihilation rate as we have discussed. 
The normalized radial wave function of $1s$ bound state of $\chi_2 \bar{\chi}_2$ is
\begin{align}
    \Psi_2^{\rm BS}(r)=\sqrt{\frac{(\mu_2\alpha)^3}{\pi}}e^{-\mu_2\alpha r}.
\end{align}
In the rest frame of the $\chi_2\bar{\chi}_2$ bound state, $p_1$ is given by 
\begin{align}
    p_1 = \frac{M}{2}\sqrt{1-\frac{4m_1^2}{M^2}},
\end{align}
where $M$ is the mass of the $1s$ bound state:
\begin{align}
    M = 2m_2\left(1-\frac{\alpha^2}{8}\right).
\end{align}
Then we obtain $\Gamma_{2\to 1}$ from eq.~\eqref{eq:gamma bound state} as
\begin{align}
    \Gamma_{2\to 1}&=\frac{m_1M|u|^2}{4\pi}\frac{(\mu_2\alpha)^3}{\pi}\sqrt{1-\frac{4m_1^2}{M^2}}\notag\\
    &=\frac{m_1m_2|u|^2}{2\pi}\frac{(\mu_2\alpha)^3}{\pi}\left(1-\frac{\alpha^2}{8}\right)\sqrt{1-\frac{4m_1^2}{M^2}}.
\end{align}
From the condition for the validity of the Born approximation, $\Gamma_{\rm 2\to 1}<\Gamma$,
we obtain the lower bound for $\Gamma/m_2$ as 
\begin{align}
    \frac{\Gamma}{m_2} \gtrsim 3.2\times 10^{-5}&\times\left(\frac{a}{10^{-7}~{\rm GeV^{-2}}}\right)\left(\frac{m_1}{1~{\rm TeV}}\right)^2\left(\frac{m_2/m_1}{1.01}\right)\left(\frac{\alpha}{0.2}\right)^3\notag\\
    &\times\left(1-\frac{\alpha^2}{8}\right)\sqrt{1-\frac{4m_1^2}{M^2}}.
\end{align}
This condition is satisfied in figure~\ref{fig:sigmav_ave_gamma}.

\begin{figure}[tbp]
    \centering
    \includegraphics[width=0.7\linewidth]{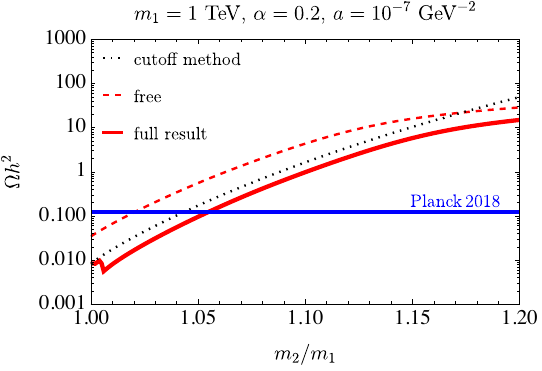}
    \\[5mm]
    \includegraphics[width=0.7\linewidth]{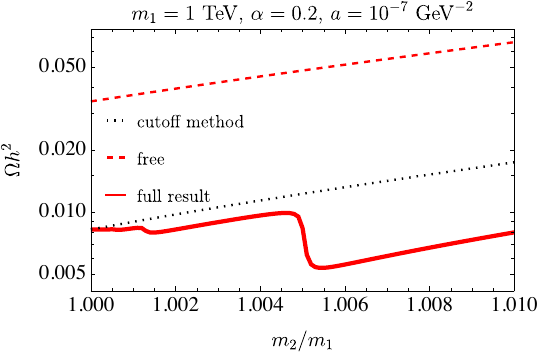}
    \caption{\emph{Top}: The DM relic abundance 
    $\Omega h^2$ as a function of the mass ratio $m_2/m_1$ in the case of the attractive Coulomb potential. The black dotted line shows the cutoff method result with $v_{\rm cut}=10^{-2}$. The red dashed line shows the free result obtained from eq.~\eqref{eq:sigvave} with $S_f(E_2,\Gamma)=1$ and $\Gamma/m_2 = 10^{-4}$. The red solid line shows the full result obtained by using eq.~\eqref{eq:sigvave} with $\Gamma/m_2 = 10^{-4}$. The blue solid line shows the central value of the observed DM energy density~\cite{Planck:2018vyg}. \emph{Bottom}: The DM relic abundance $\Omega h^2$ in the range of $1\leq m_2/m_1\leq1.01$.}
    \label{fig:abundance}
\end{figure}

Finally, we discuss the impacts of the final-state SE on the DM relic abundance $\Omega h^2$.
Figure~\ref{fig:abundance} shows the DM relic abundance 
$\Omega h^2$ of $\chi_1$ as a function of $m_2/m_1$. We solve the Boltzmann equation (\ref{eq:boltzmann}) with the thermally averaged annihilation cross sections given in eqs.~(\ref{eq:sigvave}) and (\ref{eq:sigvavecut}). 
We calculate $\Omega h^2$ in the range of $1\leq m_2/m_1\leq 1.2$. As shown in the bottom panel, the relic abundance decreases at $m_2/m_1\simeq1.0005,1.0012, 1.005$. This behavior originates from the contribution of the resonances of the bound states discussed above. Thus, the value of $\Omega h^2$ in the full result becomes smaller than the cutoff method result. As a result, the predicted mass spectra for $\chi_1$ and $\chi_2$ which reproduce the observed DM energy density deviate from the prediction of the cutoff method. If $m_2/m_1\gtrsim1.18$, the relic abundance obtained by the cutoff method becomes larger than the free result because the cutoff method does not take into account the contribution of annihilation processes with off-shell final-state particles, and underestimates the annihilation cross section.

\subsection{Attractive Hulth\'en potential}\label{sec:hulthen}
In order to discuss the effect of the bound states in more detail, we evaluate the final-state SE under the attractive Hulth\'en potential given by
\begin{align}
    V_{\rm H}(r)&=-\frac{\alpha m_* e^{-m_*r}}{1-e^{-m_*r}},
\end{align}
where $m_*$ is the screening parameter of the interaction between $\chi_2$ and $\bar{\chi}_2$. 
The potential gives a good approximation of the attractive Yukawa potential $V_{\rm Y}(r)=-\alpha e^{-m_\phi r}/r$ if we take $m_*=\pi^2m_\phi/6$, where $m_\phi$ is the mass of a mediator $\phi$. The asymptotic behaviors of $V_{\rm H}$ and $V_{\rm Y}$ at small and large $r$ are also consistent with each other.
Defining the following parameters 
\begin{align}
    \delta=\frac{m_*}{2\mu_2\alpha},\quad \hat{p}_2=\frac{p_2}{2\mu_2\alpha},\quad 
    \alpha_{\pm}=\frac{i\hat{p}_2\pm\sqrt{\delta-\hat{p}_2^2}}{\delta},
\end{align}
the solution of eq.~(\ref{eq:reducegreeneq}) with the boundary conditions (\ref{eq:reducegreenbc1}) and (\ref{eq:reducegreenbc2}) is expressed as
\begin{align}
    g_2^{\rm (H)}(r;E_2+i\Gamma)&=\frac{\Gamma(1-\alpha_+)\Gamma(1-\alpha_-)}{\Gamma(1-\alpha_+-\alpha_-)}e^{ip_2r}\,_2F_1(-\alpha_-,-\alpha_+;1-\alpha_+-\alpha_-;e^{-m_*r}),
\end{align}
where $_2F_1$ is the Gaussian hypergeometric function. Then,
\begin{align}
    \Im g_2^{\rm (H)\prime}(0,E_2+i\Gamma)
    =\Re p_2 -2\mu_2\alpha\Im[\psi(1-\alpha_+)+\psi(1-\alpha_-)].\label{eq:greenhulthen}
\end{align}
There exists a finite number of bound states under the Hulth\'en potential with small $m_*$. Its binding energy is given by
\begin{align}
    {\cal E}_n^{\rm (H)}=-\frac{\mu_2\alpha^2}{2n^2}\left(1-\frac{n^2m_*}{2\mu_2\alpha}\right)^2,\quad (n=1,2,\cdots).
\end{align}
Each bound state exists for $n$ such that $1 - n^2 m_*/2\mu_2\alpha>0$~\cite{Mitridate:2017izz},
and hence the condition of $m_*$ for the existence of at least one bound state is
\begin{align}
    m_*\leq 2\mu_2\alpha.\label{eq:criterion}
\end{align}
In the following, we compare two cases with and without a bound state.

\begin{figure}[tbp]
    \centering
    \begin{minipage}[b]{0.49\linewidth}
        \includegraphics[width=\linewidth]{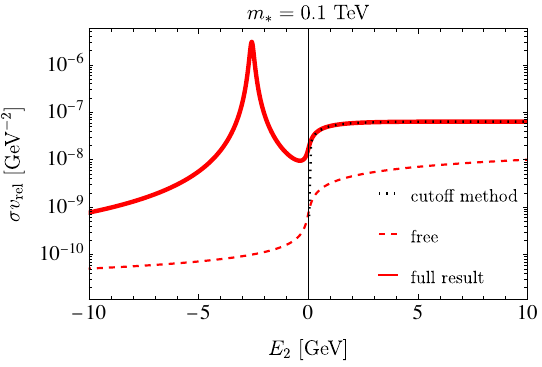}
    \end{minipage}
    \begin{minipage}[b]{0.49\linewidth}
        \includegraphics[width=\linewidth]{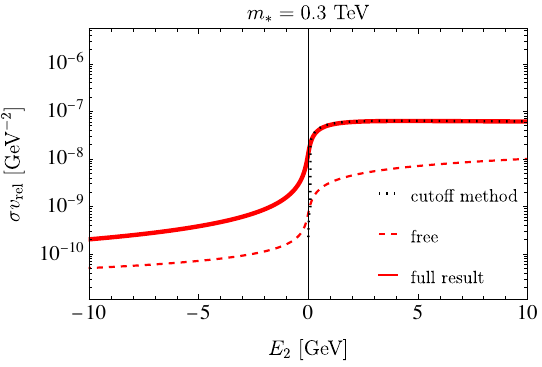}
    \end{minipage}
    \caption{Annihilation cross sections of $\chi_1\bar{\chi}_1$ as functions of $E_2$ under the attractive Hulth\'en potential. We take $m_1=1~{\rm TeV}$, $m_2/m_1=1.01$, $\alpha=0.2$, $a=10^{-7}~{\rm GeV^{-2}}$, and $\Gamma/m_2 = 10^{-4}$ in both panels. The vertical line at $E_2=0$ indicates the threshold energy of $\chi_1\bar{\chi}_1\to\chi_2\bar{\chi}_2$. The black dashed line shows the cross section including the final-state SE obtained from the cutoff method with $v_{\rm cut}=10^{-2}$. The red dashed line shows the free result with $\Gamma/m_2 = 10^{-4}$. The red solid line shows the full result of our formulation with $\Gamma/m_2 = 10^{-4}$. 
    \emph{Left}: The result for $m_*=0.1~{\rm TeV}$. \emph{Right}: The result for $m_*=0.3~{\rm TeV}$.}
    \label{fig:sigmav_hulthen_02}
\end{figure}
Figure~\ref{fig:sigmav_hulthen_02} shows $\sigma v_{\rm rel}$ as a function of $E_2$ for $m_1=1~{\rm TeV}$, $m_2/m_1=1.01$, $\alpha=0.2$, $a=10^{-7}~{\rm GeV^{-2}}$, and $\Gamma/m_2 = 10^{-4}$. 
In this case, the bound states exist if the screening parameter satisfies
$m_*\lesssim0.2~{\rm TeV}$.
In this figure, we compare the results for $m_*=0.1~{\rm TeV}$ and for $m_*=0.3~{\rm TeV}$. In the left panel, 
a peak appears at $E_2\simeq -2.6~{\rm GeV}$, which is identified as the resonance of the $1s$ bound state. 
In contrast, no resonance appears in the latter case, as shown in the right panel. As in the case of the Coulomb potential, the cross section obtained by the cutoff method vanishes below the kinematical threshold, reflecting the absence of off-shell contributions.

\begin{figure}[tbp]
    \centering
    \begin{minipage}[b]{0.49\linewidth}
        \includegraphics[width=\linewidth]{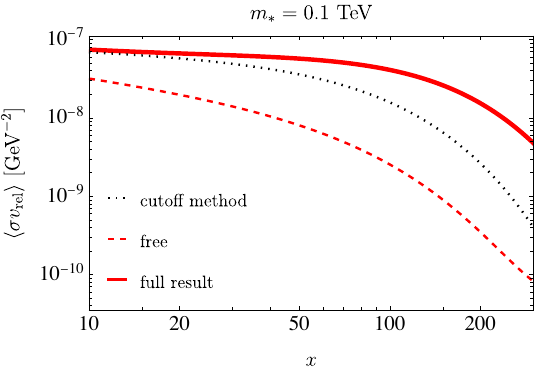}
    \end{minipage}
    \begin{minipage}[b]{0.49\linewidth}
        \includegraphics[width=\linewidth]{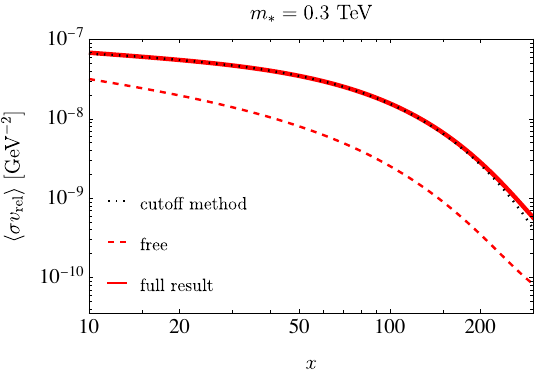}
    \end{minipage}
    \caption{Thermally averaged annihilation cross sections of $\chi_1\bar{\chi}_1$ under the attractive Hulth\'en potential as a function of $x$. In both panels, we take $m_1=1~{\rm TeV}$, $m_2/m_1=1.01$,  $\alpha=0.2$, and $a=10^{-7}~{\rm GeV^{-2}}$.
    In the left panel, we set $m_*=0.1~{\rm TeV}$, whereas in the right panel, we set $m_*=0.3~{\rm TeV}$.}
    \label{fig:sigmav_ave_hulthen}
\end{figure}
Figure~\ref{fig:sigmav_ave_hulthen} shows $\langle\sigma v_{\rm rel}\rangle$ as a function of $x=m_1/T$. Similarly to the case of the attractive Coulomb potential, 
the cutoff method underestimates the thermally averaged cross section when the bound states are present. In the left panel, the bound state resonance enhances the thermally averaged annihilation cross section by $\mathcal{O}(10)\%$ at $x = 30$.
In the absence of bound states, the deviation appears only for $x \gtrsim 200$, as shown in the right panel. This behavior originates from annihilation processes involving off-shell final-state particles, which are not included in the cutoff method. 

\begin{figure}[tbp]
    \centering
    \begin{minipage}[b]{0.49\linewidth}
        \includegraphics[width=\linewidth]{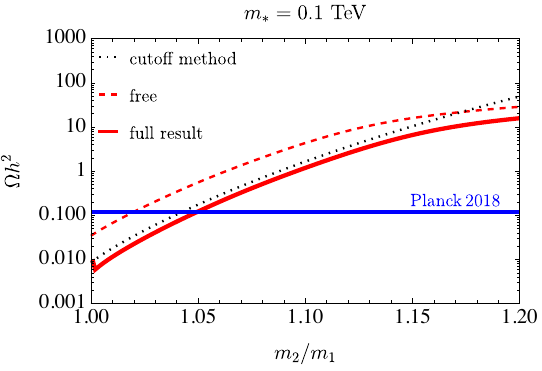}
    \end{minipage}
    \begin{minipage}[b]{0.49\linewidth}
        \includegraphics[width=\linewidth]{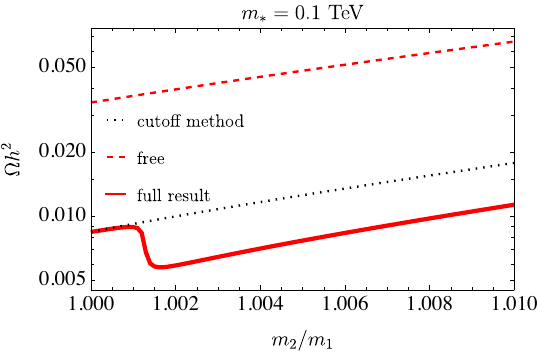}
    \end{minipage}
    ~\\[5mm]
        \includegraphics[width=0.5\linewidth]{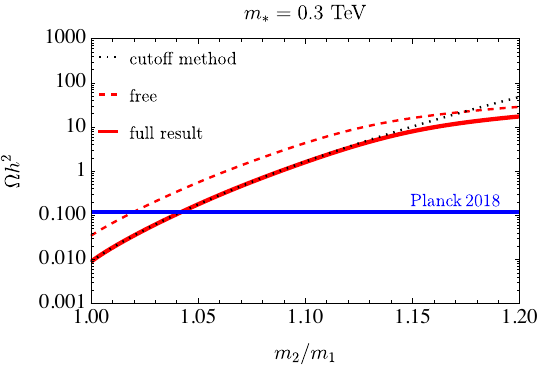}
    \caption{The relic abundance $\Omega h^2$ as a function of the mass ratio $m_2/m_1$ for the attractive Hulth\'en potential. Here, we take $m_1=1~{\rm TeV}$, $\alpha=0.2$, and $a=10^{-7}~{\rm GeV^{-2}}$. The blue horizontal line corresponds to the observed DM density~\cite{Planck:2018vyg}. 
    In the top panels, we set $m_* = 0.1$~TeV, whereas in the bottom panel we set $m_* = 0.3$~TeV. The top-right panel shows a magnified view of the top-left panel for $1 \leq m_2/m_1 \leq 1.01$.
    }
    \label{fig:abundance_hulthen}
\end{figure}
Figure~\ref{fig:abundance_hulthen} shows the DM relic abundance as a function of the mass ratio $m_2/m_1$. The top panels show the case of $m_*=0.1~{\rm TeV}$, where a bound state exists. A discrepancy between our formulation and the cutoff method appears for $m_2/m_1 \gtrsim 1.001$ due to the bound-state resonance, similar to the case of the Coulomb potential. In contrast, for $m_*=0.3~{\rm TeV}$ shown in the bottom panel, the prediction of our method does not deviate from that of the cutoff method for $m_2/m_1\lesssim1.15$ because of the absence of bound states. The deviation observed for $m_2/m_1 \gtrsim 1.15$ is due to the contribution from annihilation processes with off-shell final-state particles.

\section{Conclusion and discussion}\label{sec:conclusion}
We have studied the final-state SE of DM annihilation into heavier unstable particles which have a long-range interaction. In order to focus on the final-state SE, we do not introduce long-range interactions between particles in the initial state. We have calculated the DM annihilation cross section in the non-relativistic regime by solving the Schr\"odinger equations which include the decay width of the annihilation products. As concrete examples, we have applied our formulation for the attractive Coulomb potential and for the Hulth\'en potential. We find that our formulation automatically accounts for annihilation processes with off-shell final-state particles for the large decay width. Furthermore, similarly to $t\bar{t}$ production near the threshold, the impact of resonances of bound states on the annihilation cross section is significant, which is relevant for the narrow decay width.  This modifies the prediction of the DM relic abundance. Thus, the Schr\"odinger equations including the decay width of the annihilation products need to be solved in order to correctly evaluate the resonant effects.

We briefly comment on the effect of bound state formation (BSF) with an emission of a mediator particle $\phi$, $\chi_1\bar{\chi}_1\to B\phi$, where $B$ is the bound state of $\chi_2\bar{\chi}_2$~\cite{Wang:2022avs}. 
If $m_\phi$ is larger than the binding energy, the effect of BSF can be evaluated perturbatively, and it arises at a higher order of $\alpha$ than the resonant process, $\chi_1 {\bar\chi}_1 \to B$.
This is the case for the scenario discussed in section~\ref{sec:hulthen}.
On the other hand, for small $m_\phi$ or massless mediators, the effect of BSF cannot be calculated using perturbation theory because the soft mediators are emitted. In this case, a resummation of emission of soft $\phi$ particles is required to evaluate the DM annihilation cross section~\cite{Baumgart:2014vma,Ovanesyan:2014fwa,Bauer:2014ula,Beneke:2022pij,Fujiwara:2025cuq}. This case is beyond the scope of this paper.

Although we have concentrated on a single component DM scenario for simplicity, our calculation method can be applicable to multi-component models. For example, in the framework of accidental composite DM~\cite{Bai:2010qg,Antipin:2015xia}, dark pion DM can annihilate into heavier unstable isospin multiplet dark pions with ${\cal O}(1$--$100)~{\rm TeV}$ masses~\cite{Abe:2024mwa}. In such cases, the annihilation products are much heavier than the electroweak gauge bosons, and the interactions behave as long-range forces. Therefore, the final-state SE is naturally induced, and further discussion is required in order to evaluate the relic abundance of the dark pion DM. 

\section*{Acknowledgements}
The work of RS is supported in part by JSPS KAKENHI Grant Numbers~23K03415, 24H02236, and 24H02244. The work of TY is supported in part by JST SPRING, Grant Number JPMJSP2138.

\bibliography{ref}
\bibliographystyle{JHEP}
\end{document}